\pdfoutput=1
\documentclass[prl,twocolumn,showpacs,amsmath,amssymb,superscriptaddress,floatfix]{revtex4}

\usepackage{graphicx}
\usepackage{dcolumn}
\usepackage{bm}
\usepackage{epsfig}

\begin{document}

\title{Giant thermoelectric effect in graphene-based topological insulators with nanopores}

\author{Po-Hao Chang}
\affiliation{Department of Physics and Astronomy, University of Delaware, Newark, DE 19716-2570, USA}
\author{Mohammad Saeed Bahramy}
\affiliation{Department of Applied Physics, University of Tokyo, Tokyo 113-8656, Japan}
\affiliation{RIKEN Center for Emergent Matter Science (CEMS), Wako, Saitama 351-0198, Japan}
\author{Naoto Nagaosa}
\affiliation{Department of Applied Physics, University of Tokyo, Tokyo 113-8656, Japan}
\affiliation{RIKEN Center for Emergent Matter Science (CEMS), Wako, Saitama 351-0198, Japan}
\author{Branislav K. Nikoli\' c}
\email{bnikolic@udel.edu}
\affiliation{Department of Physics and Astronomy, University of Delaware, Newark, DE 19716-2570, USA}
\affiliation{RIKEN Center for Emergent Matter Science (CEMS), Wako, Saitama 351-0198, Japan}

\begin{abstract}
Designing thermoelectric materials with high figure of merit $ZT=S^2 G T/\kappa$ requires fulfilling three often
irreconcilable conditions, i.e., the high electrical conductance $G$,  small thermal conductance $\kappa$  and
high Seebeck coefficient $S$. Nanostructuring is one of the promising ways to achieve this goal as it can substantially suppress
lattice contribution to $\kappa$. However, it may also unfavorably influence the electronic transport in an
uncontrollable way. Here we theoretically demonstrate that this issue can be ideally solved by fabricating graphene nanoribbons
with heavy adatoms and nanopores. These systems, acting as a two-dimensional topological insulator with robust helical edge states
carrying electrical current, yield a highly optimized power factor $S^2G$ per helical conducting channel. Concurrently, their array of
nanopores impedes the lattice thermal conduction through the bulk. Using quantum transport simulations coupled with first-principles
electronic and phononic band structure calculations, the thermoelectric figure of merit is found to reach its maximum $ZT \simeq 3$
at $T \simeq 40$ K. This paves a way to design high-$ZT$ materials by exploiting the nontrivial topology of electronic states through nanostructuring.
\end{abstract}

\pacs{73.50.Lw, 03.65.Vf, 73.20.-r, 85.80.Fi}
\maketitle

Thermoelectrics~\cite{Heremans2013,Vineis2010,Tritt2011} transform temperature gradients into electric voltage and vice versa. Although a plethora of
thermoelectric energy harvesting and cooling applications has been envisioned, their usage is presently limited by their small
efficiency. This is due to the fact that increasing thermoelectric figure of merit
\begin{equation}\label{eq:zt}
ZT=\frac{S^2GT}{\kappa_{\rm el} + \kappa_{\rm ph}},
\end{equation}
requires careful trade-off between electrical conductance $G$, the Seebeck coefficient $S$, and the thermal conductance $\kappa_{\rm el} + \kappa_{\rm ph}$.
The total thermal conductance has contributions from both electrons $\kappa_{\rm el}$ and phonons (i.e., lattice vibrations) $\kappa_{\rm ph}$. $ZT$ quantifies the maximum efficiency of a thermoelectric cycle conversion in the linear-response regime where a small voltage  \mbox{$\Delta V = -S \Delta T$}
exactly cancels the current induced by the small temperature difference \mbox{$\Delta T=T_H-T_C$} at average operating temperature $T=(T_H+T_C)/2$.
The values approaching $ZT \rightarrow \infty$ would ensure Carnot efficiency as the theoretical limit for a heat engine
operating between a hot $T_H$ and a cold $T_C$ temperature. However, $ZT$ of realistic devices is limited by irreversible energy losses
via Joule heat and thermal conduction, so that a pragmatic goal is to achieve \mbox{$ZT \gtrsim 3$} with low parasitic losses and stability over a broad temperature range~\cite{Vineis2010,Tritt2011}.

The major directions to increase $ZT$ have been focused on either materials with high power factor $S^2G$,
such as doped narrow-gap semiconductors; or on minimizing $\kappa_{\rm ph}$ by enhanced phonon scattering in different frequency ranges, such as through nanostructuring~\cite{Heremans2013,Vineis2010}. Although nanostructuring has progressed rapidly over the past decade~\cite{Heremans2013,Vineis2010}, it typically affects bulk electronic states of conventional materials in unfavorable way for thermoelectricity. Thus, the recently discovered topological insulator (TI) materials~\cite{Ando2013,Hasan2010} are of particular interest. The key ingredient in this new class of materials is strong spin-orbit coupling (SOC) which opens an energy gap $E_G$ in the bulk and generates conducting edge (in two dimensions) or surface (in three-dimensions) electron states robust against backscattering off nonmagnetic disorder~\cite{Hasan2010}. Interestingly,  Bi$_2$Te$_3$ as one of the prime examples of 3D TIs~\cite{Ando2013} is well-known to be one of the best bulk thermoelectrics with $ZT \simeq 1$~\cite{Tritt2011}. Recent efforts have demonstrated how using nanocomposites of bulk and thin film Bi$_2$Te$_3$ can lead to $ZT \simeq 2.5$~\cite{Heremans2013,Vineis2010,Tritt2011}. However, none of these findings relies on the topological surface states whose contribution to
$S$ and $G$ would be insensitive to disorder introduced to suppress $\kappa_\mathrm{ph}$. The very recent attempts to design thermoelectrics based on three-dimensional~\cite{Ghaemi2010,Tretiakov2010,Tretiakov2011,Huber2012} (3D) or two-dimensional (2D)~\cite{Takahashi2010,Shevtsov2012} TIs are mostly qualitative and lack information on their phonons. In addition, many candidate 3D TI materials are unintentionally doped in the bulk which makes difficult to reach the topological transport regime where electrons, behaving as massless Dirac fermions, propagate exclusively on their surfaces~\cite{Huber2012,Ando2013}.

\begin{figure*}
\includegraphics[scale=0.43,angle=0]{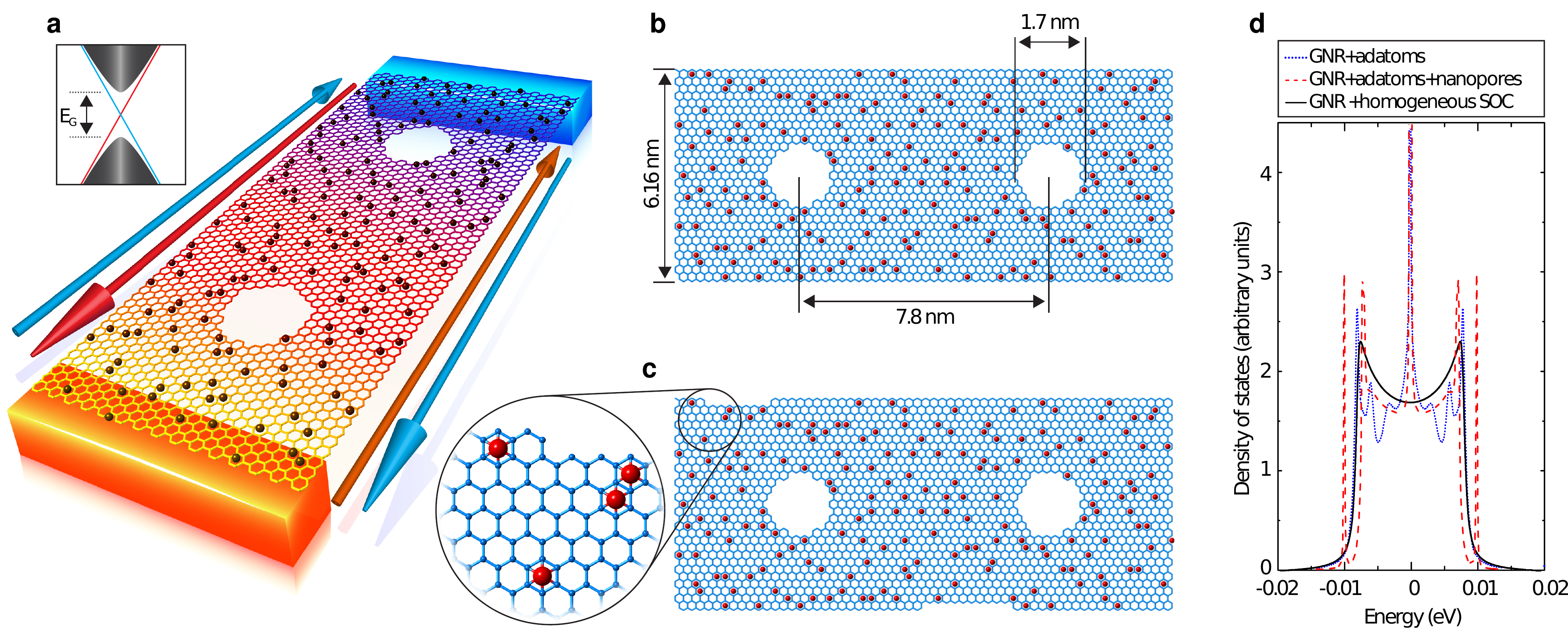}
\caption{{\bf Schematic of 2D TI thermoelectric based on graphene nanoribbons with heavy adatoms and nanopores}. (a) The central region, consisting of GNR (of width \mbox{$W=6.16$ nm} and length \mbox{$L \approx 1.35$ $\mu$m} used in our calculations) with indium adatoms and (253 along chosen $L$) nanopores, is attached to hot and cold macroscopic reservoirs. The adatoms shown in the center of randomly chosen $n_\mathrm{ad} \approx 19$\% hexagons locally enhance SOC within those hexagons which creates an energy gap $E_G$ and a pair of helical quantum states on each edge channeling flow of electrons of opposite spin in opposite directions. We consider GNRs with: (b) perfect edge of zigzag type; or (c) disordered edge created by removing one or two edge carbon atoms in each GNR supercell. (d) The total density of states, where the gap \mbox{$E_G \approx 17.3$ meV} around the Dirac point at zero energy is filled by contribution from the helical edge states, is insensitive to the randomness of adatom configuration or spatial inhomogeneities.}
\label{fig:fig1}
\end{figure*}

In this Article we present a theoretical design of a high-$ZT$ system, using graphene-based 2D TI nanowire depicted in Fig.~\ref{fig:fig1},
for which we accurately obtain all quantities in Eq.~\eqref{eq:zt} via the nonequilibrium Green function (NEGF) methodology~\cite{Stefanucci2013,Nikolic2012}  combined with first-principles calculations of both electronic and phononic band structure. Thus far, 2D TIs have been realized
experimentally using cumbersome-to-grow HgTe/CdTe~\cite{Konig2007}, or somewhat more accessible InAs/GaSb~\cite{Knez2011}, quantum wells.
A much simpler system---graphene with randomly distributed heavy adatoms---has been conjectured recently via first-principles
studies~\cite{Weeks2011,Hu2012}. For example, among many possible heavy adatoms indium and thallium favor high-symmetry position
in the center of the hexagons of honeycomb lattice of carbon atoms, while being nonmagnetic and without inducing Rashba SOC that would compete
with the emergence of the 2D TI phase~\cite{Weeks2011}. We consider two types of graphene nanoribbons (GNRs) + nanopores as
the central region of the two-terminal setup in Fig.~\ref{fig:fig1}(a), where the edge of GNR is either perfectly ordered and
chosen to be of zigzag type in Fig.~\ref{fig:fig1}(b), or lightly disordered in Fig.~\ref{fig:fig1}(c). We assume that $n_\mathrm{ad} \approx 19$\% of GNR
hexagons are randomly selected and covered by heavy adatoms of indium, which locally enhance~\cite{Weeks2011} tiny (due to
lightness of carbon atoms) intrinsic SOC coupling~\cite{Gmitra2009} already present in  graphene. This helps to increase the bulk band gap
from \mbox{$E_G/k_B \approx 0.28$ K}~\cite{Gmitra2009} in pristine graphene to \mbox{$E_G/k_B \approx 200$ K} for the chosen adatom type and $n_\mathrm{ad}$.

\begin{figure*}
\includegraphics[scale=0.4,angle=0]{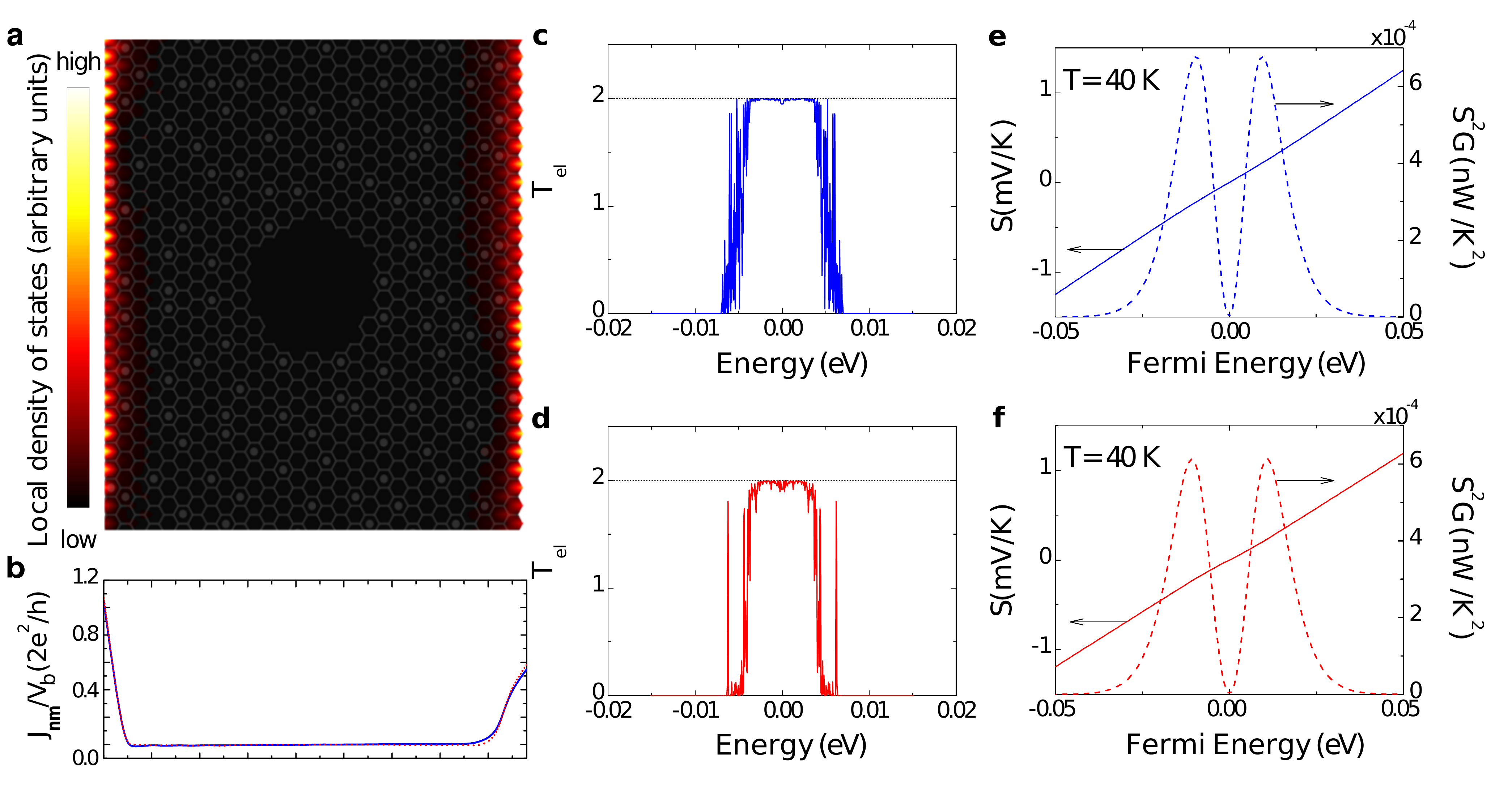}
\caption{{\bf The electronic transmission, Seebeck coefficient and power factor governed by the helical edge states}. (a) The local DOS within GNR + heavy-adatoms with a single nanopore in its interior at \mbox{$E_F=0.001$ eV}. (b) The spatial profile of local currents over the transverse cross section of system in (a). The sum of all bond currents $J_\mathbf{nm}/V_b$, describing~\cite{Zarbo2007} charge flow from site $\mathbf{m}$ to site $\mathbf{n}$ of the honeycomb lattice at bias voltage $V_b$, gives the electrical conductance $G (E_F)= e^2 T_\mathrm{el}(E_F)/h$ at zero temperature. Due to helical edge quantum states determining the local DOS and local currents, both quantities have non-zero value only around the edges. (c) and (d) The zero-bias electronic transmission function $T_\mathrm{el}(E)$ for setups in Figs.~\ref{fig:fig1}(b) and ~\ref{fig:fig1}(c), respectively. The blue and red lines in panels (b)--(d) are obtained in the presence of both heavy adatoms and nanopores, while black dotted line in panels (c) and (d) is computed after nanopores in the two setups from Fig.~\ref{fig:fig1} are removed. (e) and (f) The Seebeck coefficient at $T=40$ K computed by plugging  $T_\mathrm{el}(E)$ from panels (c) and (d) into Eq.~\eqref{eq:seebeck}, respectively. Panels (e) and (f) also show the power factor $S^2G$.}
\label{fig:fig2}
\end{figure*}

Remarkably, despite completely random position of heavy adatoms, such disordered system has extremely stable 2D TI phase (which is actually stabilized by the randomness of adatom distribution~\cite{Jiang2012}) that does not require disorder averaging~\cite{Shevtsov2012} because each sample has the same bulk energy gap
\mbox{$E_G \approx 17.3$ meV}. The gap is visible in the DOS in Fig.~\ref{fig:fig1}(d) which, together with the local DOS in Fig.~\ref{fig:fig2}(a) and related spatial profile of local charge currents in Fig.~\ref{fig:fig2}(b) (both of which are confined around the GNR edges), do not contain any signatures of spatial inhomogeneities.

Due to time-reversal symmetry (TRS), the edge currents cannot be reduced by interior nonmagnetic impurities and vacancies, or by edge disorder like the one introduced in Fig.~\ref{fig:fig1}(c). This leads to quantized zero-bias electronic transmission function $T_{\rm el}(E)$ in the Landauer-B\"{u}ttiker approach to quantum transport~\cite{Stefanucci2013}, shown in Figs.~\ref{fig:fig2}(c) and ~\ref{fig:fig2}(d) for systems illustrated in Figs.~\ref{fig:fig1}(b) and ~\ref{fig:fig1}(c), respectively. The quantized value $T_\mathrm{el}(E)=2$ reflects ballistic transport of electrons through one-dimensional counter-propagating and spin-polarized edge states, often labeled as ``helical''~\cite{Hasan2010}, where TRS forces electrons of opposite spin to flow in opposite directions. The total number of spin-polarized conducting channels on both edges is four, as illustrated in Fig.~\ref{fig:fig1}(a), where electrons in two of these channels moving in the same direction give $T_\mathrm{el}(E)=2$ (which was also employed as the experimental signature of 2D TI phase in the early experiments~\cite{Konig2007}).

We note that dotted horizontal line in Figs.~\ref{fig:fig2}(c) and ~\ref{fig:fig2}(d) plots
$T_\mathrm{el}(E)=2$ within a wider range of energies \mbox{$E_F \in [-0.4 \ \mathrm{eV}, 0.4 \ \mathrm{eV}]$}
for a uniform ZGNR + heavy-adatoms, which is inherited from the underlying subband structure of GNRs
with zigzag edges (ZGNRs)~\cite{Prada2013}. However, once nanopores and/or edge disorder are introduced the quantized $T_\mathrm{el}(E)=2$
in Figs.~\ref{fig:fig2}(c) and ~\ref{fig:fig2}(d) persists only when the Fermi energy (whose position can be
controlled by the gate voltage) is within the bulk gap $E_G$. In fact, the quantization of $T_\mathrm{el}(E)$ due to protection of helical edge states by TRS occurs in a range of energies smaller than the na\"{i}vely expected \mbox{$E_F \in [-E_G/2, E_G/2]$}. This is due to the fact that the width of non-zero LDOS around edges in Fig.~\ref{fig:fig2}(a) increases as one moves away from the Dirac point (DP) at $E_F=0$, so that when states from opposite edges start to overlap a minigap is created thereby removing the crossing point in the inset of Fig.~\ref{fig:fig1}(a) and protection by TRS~\cite{Prada2013}. Although no symmetry prevents inelastic backscattering off phonons, quantized transmission is expected to be insensitive to such inelastic mechanisms to leading order~\cite{Budich2012}.

\begin{figure*}
\centerline{\includegraphics[scale=0.45,angle=0]{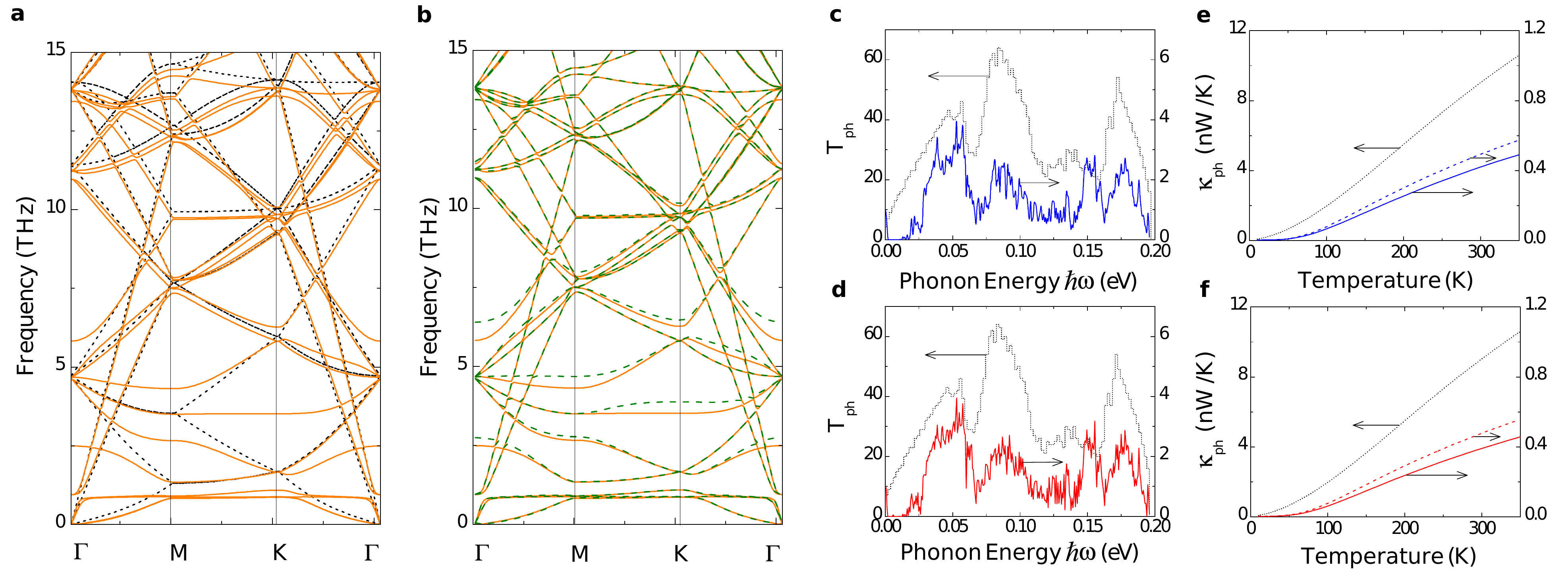}}
\caption{{\bf The phononic band structure and thermal conductance.}  Solid line in both panels (a) and (b) plots the phonon band structure for $4 \times 4$ supercell of graphene, with one indium adatom per supercell, where SOC is not included. Dotted line in panel (a) plots phonon dispersion for the same supercell of pristine graphene without any adatoms, while dashed line in panel (b) includes both adatoms and SOC. All curves are obtained from first-principles calculations using VASP~\cite{Kresse1993}. (c) and (d) The phononic transmission function $T_\mathrm{ph}(\omega)$ for setups in Fig.~\ref{fig:fig1}(b) and ~\ref{fig:fig1}(c), respectively, where solid lines are based on the dashed line dispersion in panel (b). Dotted lines in panels (c) and (d) plot quantized $T_\mathrm{ph}(\omega)$ for an infinite homogeneous ZGNR of the same width and edge shape as in Fig.~\ref{fig:fig1}, but without any adatoms, nanopores or edge disorder. (e) and (f) Solid and dotted lines plot the phononic thermal conductance, computed by plugging the values of $T_\mathrm{ph}(\omega)$ from solid and dotted lines in panels (c) and (d), respectively, into Eq.~\eqref{eq:kappaphonon}. Additional dashed line in panels (e) and (f) plots $\kappa_\mathrm{ph}$ for the same setups as in Figs.~\ref{fig:fig1}(b) and ~\ref{fig:fig1}(c), respectively, but where the presence of heavy adatoms and local SOC induced by them is neglected.}
\label{fig:fig3}
\end{figure*}

The Seebeck coefficient is obtained from the electronic transmission function as~\cite{Nikolic2012,Sivan1986}
\begin{eqnarray}
S(E_F) & = & \frac{K_1(E_F)}{eTK_0(E_F)}, \label{eq:seebeck} \\
K_n(E_F) & = & \frac{1}{h} \int\limits_{-\infty}^{\infty} dE\, T_{\rm el}(E)  (E - E_F)^n
\left(-\frac{\partial f}{\partial E} \right), \label{eq:kintegrals}
\end{eqnarray}
where \mbox{$f(E)=\{ 1 + \exp[(E-E_F)/k_BT] \}^{-1}$} is the Fermi function. The integrals in Eq.~\eqref{eq:kintegrals} also determine~\cite{Nikolic2012,Sivan1986} electronic charge \mbox{$G(E_F)=e^2K_0(E_F)$} and thermal conductance \mbox{$\kappa_{\rm el}(E_F) = \{K_2(E_F) - [K_1(E_F)]^2/K_0(E_F)\}/T$}. At low temperatures ($k_BT \ll E_F$), Eq.~\eqref{eq:seebeck} can be approximated~\cite{Sivan1986} by the so-called Mott formula
\mbox{$S(E_F) \approx (\pi^2 k_B^2 T/3e)  [d T_\mathrm{el}(E_F)/dE] [T_\mathrm{el}(E_F)]^{-1}$} which explicitly shows that large values of $S$, plotted in Figs.~\ref{fig:fig2}(e) and ~\ref{fig:fig2}(f), require steep variation of $T_{\rm el}(E)$. This occurs around \mbox{$E \simeq \pm 4.5$ meV} in Figs.~\ref{fig:fig2}(c) or \mbox{$E \simeq \pm 3.8$ meV} in Fig.~\ref{fig:fig2}(d), where the details of the steep change from $T_{\rm el}(E)=2$ to a vanishing value are also controlled by the nanopores and/or edge disorder. The electron-like (for \mbox{$E>0$}) and hole-like (for \mbox{$E<0$}) transport give contributions to $S$ of opposite sign, so that $S \equiv 0$ exactly at the DP  (as observed in the experiments on large-area graphene~\cite{Zuev2009}).

While low \mbox{$T_\mathrm{el}(E)$} also helps to increase $S$, its large values too far away from the DP are irrelevant because thermoelectric performance depends~\cite{Jeong2012} on the power factor $S^2G$. In fact, $S^2G$ shown in Figs.~\ref{fig:fig2}(e) and ~\ref{fig:fig2}(f) has exactly the same shape as the one obtained from the Mahan-Sofo (MS) mechanism~\cite{Mahan1996} based on \mbox{$T_\mathrm{el}(E)=M\delta(E)$}, which ensures  \mbox{$\kappa_\mathrm{el} \rightarrow 0$} and also sets putative~\cite{Jeong2012,Kim2009} upper limit on \mbox{$S^2G$}. However, the peak value \mbox{$(S^2G)_\mathrm{max} \simeq 8 k_B^2/h$} per spin-polarized conducting channel in Figs.~\ref{fig:fig2}(e) and ~\ref{fig:fig2}(f) is larger than the corresponding \mbox{$(S^2G)^\mathrm{MS}_\mathrm{max} \approx 5.76  k_B^2/h$} obtained~\cite{Kim2009} in the MS mechanism. Thus, 2D TI nanowires utilize their two spin-polarized edge conducting channels most efficiently for thermoelectricity, which can be traced back to the (approximate) boxcar functional shape~\cite{Jeong2012} of $T_\mathrm{el}(E)$ in Figs.~\ref{fig:fig2}(c) and ~\ref{fig:fig2}(d). Although only two available conducting channels make the total power factor of a single graphene-based 2D TI nanowire minuscule compared to bulk 3D materials~\cite{Kim2009}, quantities shown in Fig.~\ref{fig:fig2} do not scale with the wire width so that one can substantially increase the total $S^2G$ by patterning very large number of very narrow GNRs connected in parallel.

The phononic band structure in Fig.~\ref{fig:fig3}(b) for graphene supercell with indium adatoms and SOC switched on is computed from first-principles
using VASP simulation package~\cite{Kresse1993}. This serves as in input for quantum transport calculation of the phononic transmission function $T_\mathrm{ph}(\omega)$, plotted in Figs.~\ref{fig:fig3}(c) and ~\ref{fig:fig3}(d), and $\kappa_\mathrm{ph}$ in Figs.~\ref{fig:fig3}(e)
and ~\ref{fig:fig3}(f). They are mutually connected through the Landauer-type formula~\cite{Nikolic2012,Reggo1998}
\begin{equation}\label{eq:kappaphonon}
\kappa_{\rm ph} = \frac{\hbar^2}{2\pi k_B T^2} \int\limits_{0}^{\infty} d\omega\, \omega^2 T_{\rm ph}(\omega) \frac{ e^{\hbar\omega/k_BT}}{(e^{\hbar\omega/k_BT}-1)^2}.
\end{equation}
The very high Debye temperature ($\simeq 2100$ K) of graphene necessitates quantum transport treatment of phonon propagation captured by Eq.~\eqref{eq:kappaphonon}. However, this formula does not take into account the resistive umklapp phonon-phonon scattering. Nevertheless, such effect is expected to be irrelevant for GNRs depicted in Fig.~\ref{fig:fig1} because their width is much smaller than the mean-free path due to phonon-phonon scattering (e.g., \mbox{$\ell \simeq 600$ nm} or \mbox{$\ell \simeq 100$ nm} at room temperature for graphene freely suspended or lying on SiO$_2$ substrate, respectively~\cite{Pop2012}).

Although the thermal conductivity of freely suspended large-area graphene at room temperature is among the highest of any known material~\cite{Pop2012}, it decreases significantly when this 2D material is in contact with a substrate or confined into nanoribbons. We further reduce the ballistic value of $\kappa_\mathrm{ph}$ for GNRs, plotted as dotted line in Figs.~\ref{fig:fig3}(e) and ~\ref{fig:fig3}(f), by \emph{two orders} of magnitude via introduction of nanopores, as demonstrated by solid line in Figs.~\ref{fig:fig3}(e) and ~\ref{fig:fig3}(f). This reduction occurs in sufficiently long GNRs, $L \approx 1.35$ $\mu$m, where we check that decrease of $\kappa_\mathrm{ph}$ with increasing $L$ and the number of nanopores saturates around this length. Although nanopore arrays have been considered theoretically as a way to reduce lattice thermal conductivity of bulk materials~\cite{Tretiakov2011,Lee2008a}, graphene with its
high mechanical stability makes it actually possible to fabricate nanopore arrays by a variety of recently developed techniques~\cite{Tada2011}.

\begin{figure}
\includegraphics[scale=0.36,angle=0]{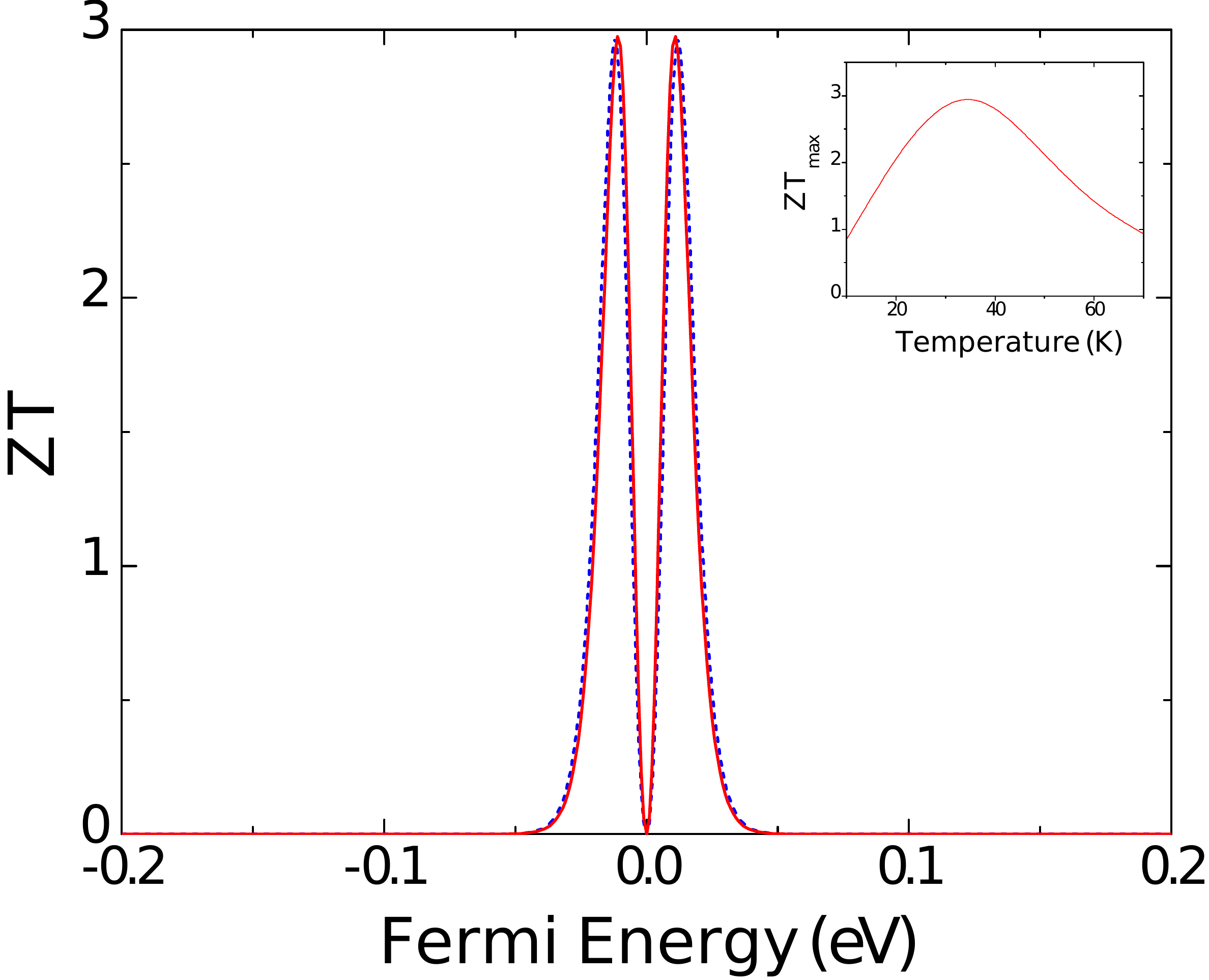}
\caption{{\bf The thermoelectric figure of merit}. Dashed line plots $ZT$ at $T=36$ K for GNR + heavy-adatoms + nanopores setup in Fig.~\ref{fig:fig1}(b), while solid line takes into account additional edge disorder for setup in Fig.~\ref{fig:fig1}(c). The inset shows dependence of the peak values of $ZT$ on temperature.}
\label{fig:fig4}
\end{figure}

The combination of electronic and phononic transport quantities shown in Figs.~\ref{fig:fig2} and ~\ref{fig:fig3}, respectively, generates  maximum $ZT_\mathrm{max} \approx 3$ for both setups Figs.~\ref{fig:fig1}(b) and ~\ref{fig:fig1}(c). Note that the main purpose of introducing light edge disorder in Fig.~\ref{fig:fig1}(c) is to highlight how GNR-based 2D TIs do not require large effort~\cite{Jia2009} to control the position of edge carbon atoms, rather
than to introduce additional phonon scattering off edge roughness~\cite{Sevincli2010,Chang2012}. The value of $ZT$ can be tuned significantly by changing the charge density (i.e., the corresponding $E_F$) via the gate voltage, where the advantage over other recent proposals~\cite{Nikolic2012,Sevincli2010,Chang2012} for thermoelectrics based on topologically trivial GNRs is insensitivity of the position of the peaks of $ZT(E_F)$ to microscopic details of the system. That is, the energy $E_F$ at which $ZT$ reaches maximum in Fig.~\ref{fig:fig4} depends only on $T$ and $E_G$ (governed by the adatom coverage and adatom type).

In summary, using quantum transport modeling combined with first-principles electronic and phononic band structure calculations we designed \emph{in silico} a high-$ZT$ thermoelectric where graphene nanoribbons with arbitrary shaped edges and nanopores in their interior are covered with heavy adatoms of indium. The adatoms provide sufficiently strong local SOC in some fraction of randomly chosen hexagons, so that such inhomogeneous SOC opens both the bulk band gap \mbox{$E_G \approx 17.3$ meV} (for $n_\mathrm{ad} \approx 19$\% of hexagons covered) and generates topologically protected helical edge states. The electronic transmission through helical edge states in the form of the (approximate) boxcar function of width $\lesssim E_G$ generates power factor $S^2G$ per conducting channel which is actually larger than the one obtained from the celebrated Mahan-Sofo mechanism using delta function transmission. This feature combined with two orders of magnitude reduction of phononic thermal conductance by the nanopores leads to the thermoelectric figure of merit for this system $ZT \simeq 3$ at low temperatures \mbox{$T \simeq 40$ K}. Since the existing bulk thermoelectric materials are very inefficient at low operating temperatures $T \sim 10$ K where they give $ZT \lesssim 0.01$~\cite{Vineis2010,Tritt2011}, TI-based high-$ZT$ thermoelectrics at low temperatures designed by our study are very attractive for applications like radioisotope thermoelectric generators on spacecrafts or cooling of electronic satellite components.

Although bulk materials are deemed necessary for large-scale power generation~\cite{Heremans2013}, GNRs underlying our proposal are single-atom-thick and with electronic transport properties which do not scale with their width, so that very high packing density~\cite{Kim2009} of GNRs connected in parallel is possible within a 3D volume.  Other choices~\cite{Hu2012} for heavy adatoms include Os, Ir and Cu-Os or Cu-Ir dimers, which could generate larger gap $E_G \gtrsim 0.2$ eV using smaller adatom coverage $n_\mathrm{ad} \simeq 2$\% (via different hybridization mechanism between carbon and adatom orbitals than in the case of indium or thallium), thereby making it possible to also tune the optimal operating temperature.  Interestingly, if the boxcar-shaped transmission function is preserved at finite bias voltage, the same system could optimize efficiency at finite power output in the nascent field of nonlinear thermoelectrics~\cite{Whitney2013}.

\section{Methods}
The electronic subsystem of GNR + heavy-adatoms is described by the tight-binding Hamiltonian of Kane-Mele type~\cite{Hasan2010} with a single $p_z$ orbital per site of the honeycomb lattice
\begin{eqnarray}\label{eq:km}
\hat{H} & = & - t \sum_{\langle \mathbf{m}\mathbf{n} \rangle,\sigma} \hat{c}_{\mathbf{m}\sigma}^\dagger \hat{c}_{\mathbf{n}\sigma} \nonumber \\
\mbox{} && + i\lambda_\mathrm{SO} \sum_\mathcal{P} \sum_{\langle \langle \mathbf{m}\mathbf{n} \rangle \rangle \in \mathcal{P},\sigma,\sigma^\prime} \nu_\mathbf{nm} \hat{c}_{\mathbf{m}\sigma}^\dagger \hat{s}^z_{\sigma \sigma^\prime} \hat{c}_{\mathbf{n}\sigma^\prime}.
\end{eqnarray}
Here the operator $\hat{c}_{\mathbf{m}\sigma}^\dagger$ ($\hat{c}_{\mathbf{m}\sigma}$) creates (annihilates) electron on site $\mathbf{m}$ of the lattice in spin state $\sigma$ and $\hat{s}^z$ is the Pauli matrix. The nearest-neighbor hopping \mbox{$t=2.7$ eV} in the first term in Eq.~\eqref{eq:km} sets the unit of energy scale. The spin-dependent hopping in the second term, where $\nu_\mathbf{mn} = 1$ for moving counterclockwise around the hexagon and $\nu_\mathbf{mn} = -1$ otherwise, acts between next-nearest neighbor sites of \emph{only} those hexagons $\mathcal{P}$ of the honeycomb lattice which host indium adatoms. The strength of such SOC, which can be viewed as locally enhanced version of the tiny intrinsic SOC in pristine graphene~\cite{Gmitra2009}, is parameterized by $\lambda_\mathrm{SO}$.

This minimal effective model in Eq.~\eqref{eq:km} is sufficient~\cite{Weeks2011} to fit---using \mbox{$\lambda_\mathrm{SO}=0.0037t$}---the low-energy spectrum obtained from first-principles calculations for $4 \times 4$ graphene supercell with two indium adatoms using VASP package~\cite{Kresse1993}, which gives \mbox{$E_G \approx 11.5$ meV}. The electron-core interactions are described by the projector augmented wave (PAW) method~\cite{Blochl1994,Kresse1999}, and we use Perdew-Burke-Ernzerhof (PBE)~\cite{Perdew1996} parametrization of the generalized gradient approximation (GGA) for the exchange-correlation functional. The cutoff energies for the plane wave basis set used to expand the Kohn-Sham orbitals are 500 eV for all calculations. A \mbox{$11 \times 11 \times 1$} $k$-point mesh within Monkhorst-Pack scheme is used for the Brillouin zone (BZ)  integration. Structural relaxations and total energy calculations are performed ensuring that the Hellmann-Feynman forces acting on ions are less than $0.005$ eV/\AA.

Starting from the matrix representation $\mathbf{H}$ of the Hamiltonian in Eq.~\eqref{eq:km}, we compute the electronic retarded GF~\cite{Stefanucci2013},
\mbox{$\mathbf{G}(E) = [E - \mathbf{H} - {\bm \Sigma}_L(E) - {\bm \Sigma}_R(E)]^{-1}$}, where ${\bm \Sigma}_{L,R}$ are the self-energies introduced by
the semi-infinite ideal (without disorder, adatoms or nanopores) ZGNR leads assumed to be attached to 2D TI wire in Figs.~\ref{fig:fig1}(b)
or ~\ref{fig:fig1}(c). The retarded GF and the level broadening matrices  ${\bm \Gamma}_{L,R} (E) = i[{\bm \Sigma}_{L,R}(E)-{\bm \Sigma}_{L,R}^\dagger(E)]$ allow us to obtain the electronic zero-bias transmission function, $T_{\rm el}(E) = {\rm Tr} \left\{ {\bm \Gamma}_R (E)  {\bf G}(E) {\bm \Gamma}_L (E)  {\bf G}^\dagger(E)  \right\}$, which determines electronic transport quantities through Eq.~\eqref{eq:kintegrals}.

The phononic band structure plotted in Figs.~\ref{fig:fig3}(a) and ~\ref{fig:fig3}(b) was computed via first-principles methodology using combined VASP~\cite{Kresse1993} and Phonopy packages~\cite{Togo2008}. The details of VASP calculations are the same as delineated above (except that we use
$3 \times 3 \times 1$ $k$-point mesh), but here we start from $4 \times 4$ graphene supercell hosting one indium adatom and then enlarged it to $8 \times 8$ supercell in order to capture accurately force constants between a range of neighboring carbon atoms or carbon atoms and indium adatoms. Figures~\ref{fig:fig3}(a) and ~\ref{fig:fig3}(b) demonstrate appearance of new low energy bands due to the presence of indium adatoms. Although the effect of SOC on phononic band structures can be profound for materials containing heavy elements, especially on surfaces and in thin films (as exemplified by the recent calculations on Bi$_2$Te$_3$ and Bi$_2$Te$_3$~\cite{Huang2012a}), the inclusion of SOC in Fig.~\ref{fig:fig3}(b) generates only a small difference.

To construct the empirical up to fourth-nearest neighbors force constant model, we varied and optimized the FCs to fit as closely as possible the phononic dispersions plotted in Figs.~\ref{fig:fig3}(a) and ~\ref{fig:fig3}(b). Using the FC matrix ${\bf K}$, the diagonal matrix ${\bf M}$ containing atomic masses, and self-energies ${\bm \Pi}_{L,R}$ of the semi-infinite ideal ZGNR leads [obtained using FCs extracted from the dotted line in Fig.~\ref{fig:fig3}(a)], we compute the phononic version of the retarded GF~\cite{Nikolic2012}, \mbox{${\bf D}(\omega)=[\omega^2 {\bf M} - {\bf K} - {\bm \Pi}_L(\omega) - {\bm \Pi}_R(\omega)]^{-1}$}. This, together with the level broadening matrices \mbox{${\bm \Lambda}_{L,R}(\omega)=i[{\bm \Pi}_{L,R}(\omega) - {\bm \Pi}_{L,R}^\dagger(\omega)]$}, gives the phononic transmission function, \mbox{$T_{\rm ph}(\omega) = {\rm Tr} \left\{ {\bm \Lambda}_R (\omega)  {\bf D}(\omega) {\bm \Lambda}_L (E)  {\bf D}^\dagger(\omega)  \right\}$}, which determines $\kappa_\mathrm{ph}$ through Eq.~\eqref{eq:kappaphonon}.

The significant difference between $\kappa_\mathrm{ph}$ for GNRs with nanopores but neglecting heavy adatoms  [dashed line in Figs.~\ref{fig:fig3}(e) and ~\ref{fig:fig3}(f)] and $\kappa_\mathrm{ph}$ when heavy adatoms and the corresponding SOC are included  [solid line in Figs.~\ref{fig:fig3}(e) and ~\ref{fig:fig3}(f)] confirms the necessity for the procedure delineated above. We note that the values of $\kappa_\mathrm{ph}$ [solid lines in Figs.~\ref{fig:fig3}(e) and ~\ref{fig:fig3}(f)] based on FCs extracted from the phononic band structure of bulk graphene with heavy adatoms in Fig.~\ref{fig:fig3}(b) are most likely overestimated---more precise FCs would require computationally very expensive procedure which considers a large
number of atoms confined within the nanoribbon geometry and in the presence of nanopores~\cite{Nikolic2012,Chang2012}.

\section{Acknowledgments}
P.-H.C. and B.K.N. were supported by US NSF under Grant No. ECCS 1202069. M.S.B. and N.N. were supported by Grant-in-Aids for Scientific Research (21244053) from the Ministry of Education, Culture, Sports, Science and Technology of Japan, Strategic International Cooperative Program (Joint Research Type) from Japan Science and Technology Agency, and also by Funding Program for World-Leading Innovative R\&D on Science and Technology (FIRST Program).




\end{document}